# Mass spectrum of Mesons via the WKB approximation method.

## Ekwevugbe. Omugbe[1], Omosede. E. Osafile[1] and Michael. C. Onyeaju[2]

[1]Department of physics, Federal University of Petroleum Resources, Effurun, Delta State, Nigeria.

[2]Theoretical Physics Group, Department of Physics, University of Port Harcourt, Port Harcourt, Nigeria.

Corresponding author email: omugbeekwevugbe@gmail.com

In this paper, we demonstrated that the multiple turning point problems within the framework of the Wentzel-Kramers-Brillouin (WKB) approximation method can be reduced to two turning point one for a non-symmetric potential function by using an appropriate Pekeris-type approximation scheme. We solved the Schrödinger equation with the Killingbeck potential plus an inversely quadratic potential (KPIQP) function. The special cases of the modeled potential are discussed. We obtained the energy eigenvalues and the mass spectra of the heavy $(Q\bar{Q})$ and heavy-light $(Q\bar{q})$ mesons systems. The results in this present work are in good agreement with the results obtained by other analytical methods and available experimental data in the literature.

1. Introduction

In the Quark model, a meson is a hadronic sub-particle which consists of the quark and its anti-quark and the reduced mass is dominated by the light quark mass [1]. The meson system is mediated by the strong interactions which are described by the theory of quantum chromo-dynamics (QCD) [2]. The heavier mesons also known as quarkonia $(Q\bar{Q})$ are the constituents of the heavy quarks such as the bottom (b) and charmed (c) and are considered as non-relativistic bound systems described by the Schrodinger wave equation (SE) [1, 2]. However, the case is different for the Light (uds) quark interaction systems $(q\bar{q})$ which are described by relativistic equations [1, 2]. Also, the heavy-light meson $(Q\bar{q})$ system bound states have been studied using both the relativistic and non-relavistic quark models [1, 2]. The bound state solutions to the wave equations under the quark-antiquark interaction potential functions such as the Killingbeck or the Cornell potentials have attracted much research interest in atomic and high energy physics [3-26]. The Killingbeck potential comprises the sum of the Cornell plus the Harmonic oscillator potentials, while the Cornell potential is the sum of the Coulomb plus linear potentials. The Cornell potential and its extended forms have been extensively solved with the SE [3-22], Semi-relativistic equation [23], and also with the relativistic equation [24-26]. The solution of the wave equation with some potentials are exactly solvable for $l=0$, whereas other potentials are insoluble and nontrivial for any arbitrary angular momentum quantum number $(l \neq 0)$. In this case, numerical methods and approximate analytical techniques are required to obtain the solutions of the quantum system of choice [27, 28]. Owing to the non-trivial mathematical properties of the quark-antiquark interaction potentials, several analytical techniques and approximate methods have been used to calculate the energy and mass spectra. The standard methods used in the literature are the Nikoforov Uvarov (NU) method [12-17, 29], Asymptotic iterative method (AIM) [30], the Laplace transformation method (LTM) [18], Artificial neural network method (ANN)[19] and the Analytical exact iterative method (AEIM) [11]. The authors in [12-17] obtained the mass spectrum of the quark-antiquark interaction system using an appropriate Pekeris-type approximation scheme to deal with the orbital centrifugal energy barrier. In this present work, motivated by their approach, the same Pekeris-type approximation scheme is extended to the WKB formalism for the first time to the best of our knowledge. It is well known that the multiple turning points problems within the framework of the WKB approximation method can be reduced to the standard two turning points problem for symmetric potential functions for example, the isotropic harmonic oscillator and the molecular pseudoharmonic potential with the proper coordinate transformation [31 -33]. This approach allows us to obtain the exact bound state solutions of the SE for any arbitrary $l$ quantum numbers in the physical axis ($0 < r_1 < r_2$), where $r_1, r_2$ are the classical turning points.

In this present work, we studied the non-relativistic quark model under the interaction of the non-symmetric KPIQP. The main focus of this paper is to obtain the mass spectra of the heavy meson such as charmonium $(c\bar{c})$, bottomonium $(b\bar{b})$, bottom-charmed $(b\bar{c})$, and the heavy-light meson such as the charmed-strange $(c\bar{s})$ system via the WKB approximation method.

The KPIQP has the form [15, 30] of

$$V(r) = Ar^2 + Br - \frac{C}{r} + \frac{D}{r^2}, \qquad (1)$$

where $A, B, C$ and $D$ are constant potential parameters.

If we set $A = B = D = 0$, the KPIQP reduces to the Coulomb potential used in the description of the hydrogenic atom.

$$V(r) = -\frac{C}{r} \qquad (2)$$

The KPIQP reduces to the Cornell potential if we set the constants $(A = D = 0)$

$$V(r) = Br - \frac{C}{r}, \quad B, C > o. \tag{3}$$

where C is a coupling constant and B is a linear confinement parameter.

The SE spectrum generated by the Cornell potential can be used in the investigation of the masses and decay widths of charmonium states [8, 19]. The coulombic term arises from one gluon exchange between the quark and its anti-quark and dominates at short distances [25]. While the linear term which is supported by lattice QCD measurements, dominates at large distances [19]. The addition of an inversely quadratic potential (IQP) and the harmonic oscillator (HO) term to the Cornell potential improves the behavior of the potential in the region $r \to 0$. Furthermore, it leads to improved results as compared to the Cornell potential [30]

The remaining part of the paper is organized as follows. In section two, the synopsis of the WKB approximation formalism is presented. Section three contains the analytical solution of the bound states of SE generated by the KPIQP. The special cases of the obtained energy eigenvalue are discussed. In section four, we present the numerical results of the masses of charmonium($c\bar{c}$), bottomonium $(b\bar{b})$, bottom-charmed $(b\bar{c})$, and charmed-strange $(c\bar{s})$ mesons. Furthermore, we compared the WKB mass spectra with the ones obtained by other analytical methods and available experimental data. The paper is concluded in section five.

## 2. WKB Approximation Formalism

The WKB approximation method is an effective tool initially proposed to find approximate solutions to the one dimensional time independent SE in the limiting case of large quantum numbers. It is used to obtain the finite wave function and energy eigenvalues of potentials of interest [20, 31-43]. The method can be used to study quantum tunneling rates in a potential barrier, resonance behavior in a continuum, the exponential decay of an unstable system [43] and the quasi-normal nodes of a black hole and quantum cosmology [44]. The method fails at the classical turning point where the momentum vanishes. This failure is one of the major problem associated with the method. The difficulty can be circumvented using the connection formula [42, 43]. The method accuracy varies markedly for the ground and other low lying states depending on the potential function [39]. Also, the leading order WKB approximation scheme does not yield an exact eigenvalue of the radial SE [31]. To overcome this problem, the orbital centrifugal barrier term $l(l+1)$ in the radial SE has to be replaced with the term $\left(l+\frac{1}{2}\right)^2$. This modification is known as the Langer correction [40]. Sergeenko [38], stated that the Langer correction regularizes the WKB wave function at the origin and ensures the correct asymptotic behaviour at large radial quantum numbers. Also, the centrifugal barrier contribution of the effective potential does not vanish for the s-wave case ($l = 0$) which makes the SE non-trivial for some potential functions [31, 45], hence the need to use an appropriate approximation scheme. The three-dimensional time-independent SE with a reduced mass $\mu$ and wave-function $\psi(r,\theta,\phi)$ is given as

$$\frac{\hbar^2}{2\mu}\left[\frac{1}{r^2}\frac{\partial}{\partial r}\left(r^2\frac{\partial}{\partial r}\right)+\frac{1}{r^2\sin\theta}\frac{\partial}{\partial\theta}\left(\sin\theta\frac{\partial}{\partial\theta}\right)+\frac{1}{r^2\sin^2\theta}\frac{\partial^2}{\partial\phi^2}\right]\psi(r,\theta,\phi) \quad (4)$$
$$+V(r)\psi(r,\theta,\phi) = E\psi(r,\theta,\phi).$$

With the help of the method of separation of variables, we can obtain the radial SE by using the transformation $\psi(r,\theta,\phi) = \frac{R(r)Y(\theta,\phi)}{r}$.

$$\frac{d^2R(r)}{dr^2}+\frac{2\mu}{\hbar^2}\left[E-V(r)-\frac{\left(l+\frac{1}{2}\right)^2\hbar^2}{2\mu r^2}\right]R(r)=0, \quad (5)$$

where the effective potential is given as

$$V_{eff}(r) = V(r) + \frac{\left(l+\frac{1}{2}\right)^2\hbar^2}{2\mu r^2} \quad (6)$$

We can rewrite Eq. (5) as

$$\left[\left(-i\hbar\frac{d}{dr}\right)^2\right]R(r) = 2\mu\left[E-V(r)-\frac{\left(l+\frac{1}{2}\right)^2\hbar^2}{2\mu r^2}\right]R(r). \quad (7)$$

From Eq. (7) the classical momentum is given as

$$P(r) = \left\{2\mu\left[E-V(r)-\frac{\left(l+\frac{1}{2}\right)^2\hbar^2}{2\mu r^2}\right]\right\}^{\frac{1}{2}}. \quad (8)$$

The standard WKB quantization condition [38, 39] for two turning points ($r_1, r_2$) problem is given as

$$\int_{r_1}^{r_2}P(r)dr = \pi\hbar\left(n+\frac{1}{2}\right). \quad r_1 < r < r_2 \quad n = 0, 1, 2 \cdots \quad (9)$$

The turning points are gotten from Eq. (8) by setting $P(r) = 0$.

The semi-classical wave function in the leading $\hbar$ approximation has the form

$$\psi^{WKB}(r) = \frac{N}{\sqrt{P(r)}} exp\left[\pm \frac{i}{\hbar}\int P(r)dr\right]. \tag{10}$$

## 3. Solution of the radial Schrodinger equation

In this section, we obtain the bound state solution of the SE by substituting Eq. (1) into the WKB standard quantization condition given by Eq. (9)

$$\int_{r_1}^{r_2}\left\{2\mu\left[E - Ar^2 - Br + \frac{C}{r} - \frac{D}{r^2} - \frac{\left(l+\frac{1}{2}\right)^2 \hbar^2}{2\mu r^2}\right]\right\}^{\frac{1}{2}} dr = \pi\hbar\left(n+\frac{1}{2}\right), \quad r_1 < r < r_2 \quad n = 0, 1, 2 \ldots \tag{11}$$

In this case, the classical momentum is given as

$$P(r) = \left\{2\mu\left[E - Ar^2 - Br + \frac{C}{r} - \frac{D}{r^2} - \frac{\left(l+\frac{1}{2}\right)^2 \hbar^2}{2\mu r^2}\right]\right\}^{\frac{1}{2}} \tag{12}$$

Equation (12) will produce four turning points $r_1$, $r_2$, $r_3$ and $r_4$ of which two of the points must lie in the real axis $(r > 0)$ where $P(r) = 0$. The multiple turning point problem with the potential in Eq (1) is beset with some mathematical challenges. (i) The turning points of the polynomial equation in Eq. (12) would have to be determined by some means of algebra. (ii) The evaluation of the integral in Eq. (11) is not so easy to solve analytically even without the centrifugal barrier term.

To obtain the analytical solution of the WKB quantization integral in Eq. (11), we used the Pekeris-type approximation scheme by setting $r = \frac{1}{y}$.

Changing the variable from $r$ to $y$, we obtained

$$-\sqrt{2\mu}\int_{y_1}^{y_2}\frac{dy}{y^2}\sqrt{E - \frac{A}{y^2} - \frac{B}{y} + Cy - Dy^2 - \frac{\left(l+\frac{1}{2}\right)^2 \hbar^2 y^2}{2\mu}} =$$

$$\pi\hbar\left(n+\frac{1}{2}\right), \quad y_1 < y < y_2 \quad n = 0, 1, 2 \tag{13}$$

Next, we expanded the terms $\frac{A}{y^2}$ and $\frac{B}{y}$ in power series form to the second-order around $r_0 \left(\delta = \frac{1}{r_0}\right)$ which is assumed to be the characteristic radius of a meson [12-17].

If we let $x = y - \delta$ and expand around $x = 0$ we obtained

$$\frac{A}{y^2} = \frac{A}{(x+\delta)^2} = \frac{A}{\delta^2}\left(1 + \frac{x}{\delta}\right)^2 \approx \frac{A}{\delta^2}\left(1 - \frac{2x}{\delta} + \frac{3x^2}{\delta^2}\right) \approx A\left(\frac{6}{\delta^2} - \frac{8y}{\delta^3} + \frac{3y^2}{\delta^4}\right). \tag{14}$$

In the same vein, we obtained the expansion for $\frac{B}{y}$ as

$$\frac{B}{y} \approx B\left(\frac{3}{\delta} - \frac{3y}{\delta^2} + \frac{y^2}{\delta^3}\right). \tag{15}$$

On substituting Eqs. (14) and (15) into (13) we obtained

$$-\sqrt{2\mu}\int_{y_1}^{y_2} \frac{dy}{y^2} \sqrt{-T + My - Ny^2} = \pi\hbar\left(n + \frac{1}{2}\right). \; y_1 < y < y_2 \quad n = 0, 1, 2 \dots, \tag{16}$$

where

$$N = \left(D + \frac{3A}{\delta^4} + \frac{B}{\delta^3} + \frac{\left(l+\frac{1}{2}\right)^2 \hbar^2}{2\mu}\right) \tag{17}$$

$$M = \left(C + \frac{8A}{\delta^3} + \frac{3B}{\delta^2}\right) \tag{18}$$

$$-T = \left(E - \frac{6A}{\delta^2} - \frac{3B}{\delta}\right) \tag{19}$$

Equation (16) can further be simplified as

$$-\sqrt{2\mu N}\int_{y_1}^{y_2} \frac{dy}{y^2} \sqrt{-C + By - y^2} = \pi\hbar\left(n + \frac{1}{2}\right). \; y_1 < y < y_2 \quad n = 0, 1, 2 \dots, \tag{20}$$

where

$$B = \frac{M}{N} \tag{21}$$

$$C = \frac{T}{N} \tag{22}$$

Furthermore, we can write Eq. (20) in a more compact or standard form:

$$-\sqrt{2\mu N}\int_{y_1}^{y_2}\frac{dy}{y^2}\sqrt{(y_2-y)(y-y_1)}=\pi\hbar\left(n+\frac{1}{2}\right),\quad 0<y_1<y_2 \tag{23}$$

Where $y_1, y_2$ are real turning points obtained by solving the quadratic equation in the squared root of the integrand of Eq. (20).

$$y_1=\frac{B-\sqrt{B^2-4C}}{2} \tag{24}$$

$$y_2=\frac{B+\sqrt{B^2-4C}}{2} \tag{25}$$

Performing the integration of (23), we obtained the solution

$$\int_{y_1}^{y_2}\frac{dy}{y^2}\sqrt{(y_2-y)(y-y_1)}=\frac{\pi\left[\sqrt{y_2 y_1}-\frac{1}{2}(y_2+y_1)\right]}{\sqrt{y_2 y_1}},\quad 0<y_1<y_2 \tag{26}$$

By comparing Eqs. (23) and (26), implies that

$$\frac{\sqrt{2\mu N}\left[\frac{1}{2}(y_2+y_1)-\sqrt{y_2 y_1}\right]}{\sqrt{y_2 y_1}}=\hbar\left(n+\frac{1}{2}\right). \tag{27}$$

Furthermore, with the help of Eqs. (24) and (25), we can write the sum and products of the turning points as

$$(y_2+y_1)=B=\frac{M}{N} \tag{28}$$

$$y_2 y_1=C=\frac{T}{N} \tag{29}$$

substituting Eqs. (28) and (29) into (27), we obtained the expression

$$T=\frac{\mu}{2}\left[\frac{M}{\hbar\left(n+\frac{1}{2}\right)+\sqrt{2\mu N}}\right]^2 \tag{30}$$

Finally, using the respective notations $T, M,$ and $N$ given in Eqs. (17-19), we obtained the energy eigenvalue expression for the KPIQP as

$$E_{nl}=-\frac{2\mu}{\hbar^2}\left[\frac{\left(\frac{8A}{\delta^3}+C+\frac{3B}{\delta^2}\right)}{(2n+1)+\sqrt{\frac{8\mu}{\hbar^2}\left(\frac{3A}{\delta^4}+\frac{B}{\delta^3}+D+\frac{\left(l+\frac{1}{2}\right)^2\hbar^2}{2\mu}\right)}}\right]^2+\frac{6A}{\delta^2}+\frac{3B}{\delta} \tag{31}$$

## 4. Discussion

In this section, we present some special cases of the energy eigenvalues of the KPIQP and also obtain the mass spectra of quarkonia and the heavy-light meson system.

If we set $A = B = D = 0$ and $C = Ze^2$ in Eq. (31) we immediately retrieved the Coulomb's energy eigenvalue equation given as

$$E_{nl}^{CP} = -\frac{\mu Z^2 e^4}{2\hbar^2 n_p^2}, \quad n_p = 1, 2, 3 \tag{32}$$

where $Z$ is the atomic number, $e$ is the electronic charge and $n_p$ is a principal quantum number with the notation $n_p = n + l + 1$

Setting $A = B = 0$, $D \neq 0$, $C \neq 0$, Eq. (1) reduces to the molecular Kratzer-type [46-59] potential and the energy eigenvalue equation has the form of

$$E_{nl} = -\frac{2\mu}{\hbar^2}\left[\frac{C}{(2n+1)+\sqrt{\frac{8\mu}{\hbar^2}\left(D+\frac{\left(l+\frac{1}{2}\right)^2 \hbar^2}{2\mu}\right)}}\right]^2 \tag{33}$$

If we let $A = D = 0$, then Eq. (31) reduces to the approximate energy expression of the Cornell potential,

$$E_{nl} = -\frac{2\mu_{q\bar{q}}}{\hbar^2}\left[\frac{\left(C+\frac{3B}{\delta^2}\right)}{(2n+1)+\sqrt{\frac{8\mu}{\hbar^2}\left(\frac{B}{\delta^3}+\frac{\left(l+\frac{1}{2}\right)^2 \hbar^2}{2\mu}\right)}}\right]^2 + \frac{3B}{\delta} \tag{34}$$

In order to compute the mass spectra of meson systems, we use the meson mass relation [12]

$$M_{nl} = m_q + m_{\bar{q}} + E_{nl} \tag{35}$$

Where $m_q$ and $m_{\bar{q}}$ are the respective quark and anti-quark masses.

By substituting Eq.(34) into (35), the mass spectrum of the meson systems for any arbitrary radial and angular momentum quantum numbers becomes

$$M_{nl} = m_q + m_{\bar{q}} - \frac{2\mu_{q\bar{q}}}{\hbar^2}\left[\frac{\left(C+\frac{3B}{\delta^2}\right)}{(2n+1)+\sqrt{\frac{8\mu}{\hbar^2}\left(\frac{B}{\delta^3}+\frac{\left(l+\frac{1}{2}\right)^2 \hbar^2}{2\mu}\right)}}\right]^2 + \frac{3B}{\delta} \tag{36}$$

Where $\mu_{q\bar{q}}$ is the reduce mass of the quark–antiquark systems given by the relation

$$\mu_{q\bar{q}} = \frac{m_q m_{\bar{q}}}{m_q + m_{\bar{q}}} \tag{37}$$

With the numerical support of the MAPLE package, we generated the masses of $(Q\bar{Q})$ and $(Q\bar{q})$ mesons systems by fitting Eq. (36) with experimental data given in Tables 1-4. For the charmonium meson, we substituted the experimental data for the 1S, 2S, 3S states into Eq. (36) and obtained the free parameters $B, C$ and $\delta$ by solving three algebraic equations. For the bottomonium meson, we substituted the value of $B$ obtained from the charmonium fit, including the substitution of the experimental values for the 1S and 2S states into Eq. (36) to obtain the parameters $C$ and $\delta$. Also, for the bottom-charmed meson, the values of $B, C$ gotten from the charmonium fit with the 1S state experimental data are substituted into Eq. (36) to obtain the parameter $\delta$. Finally, the free parameters for the Charmed-strange meson are obtained by inserting $B$ and the experimental data for the 1S and 2S states into Eq. (36). This enabled us to determine the $C$ and $\delta$ parameters from two algebraic equations. The masses for other excited states are obtained and compared with the results obtained by other analytical methods including available experimental data. Generally, the mass spectra increase as the adjacent quantum levels increases. These trends are in consonance with the results reported in the literature [11, 15, 18, 19, 30], including available experimental data [60-62]. Also, the results for the Charmed-strange meson tabulated in Table 4, are in excellent agreement with the ones obtained in Refs. [15, 30] and experimental data such as the excited 1D-state [62]. Furthermore, in Table 3, the mass spectrum of the bottom-charmed meson are very close to the ones obtained with the artificial neural network method [19] and also to the 2S state experimental data [60] indicating an improvement compared to the other methods.

Table1. Charmonium $(c\bar{c})$ spectra in $GeV$ ($m_c = 1.209 GeV$, $B = 0.202\ GeV^2$, $C = 1.213$, $\delta = 0.236\ GeV$).

| State | Present work | AEIM[11] | NU [15] | AIM[30] | LTM[18] | ANN [19] | Exp. [60] |
|---|---|---|---|---|---|---|---|
| 1S | 3.098 | 3.095481 | 3.095 | 3.096 | 3.0963 | 3.098 | 3.097 |
| 2S | 3.689 | 3.567354 | 3.685 | 3.686 | 3.5681 | 3.688 | 3.686 |
| 3S | 4.041 | 4.039226 | 4.040 | 4.275 | 4.0400 | 4.029 | 4.039 |
| 4S | 4.266 | 4.511098 | 4.262 | 4.865 | 4.5119 | | |
| 1P | 3.262 | 3.567735 | 3.258 | 3.214 | 3.5687 | 3.516 | 3.511 |
| 2P | 3.784 | 4.039607 | 3.779 | 3.773 | 4.0406 | 3.925 | 3.927 |
| 1D | 3.515 | 4.039683 | 3.510 | 3.412 | 4.0407 | 3.779 | 3.770 |

Table2. Bottomonium $(b\bar{b})$ spectra in $GeV$ ($m_b = 4.823\ GeV$, $B = 0.202\ GeV^2$, $C = 1.664$ $\delta = 0.361\ GeV$).

| State | Present work | AEIM[11] | NU [15] | AIM[30] | LTM[18] | ANN [19] | Exp. [60] |
|---|---|---|---|---|---|---|---|
| 1S | 9.461 | 9.74473 | 9.460 | 9.460 | 9.745 | 9.460 | 9.460 |
| 2S | 10.023 | 10.02315 | 10.022 | 10.023 | 10.023 | 10.026 | 10.023 |
| 3S | 10.365 | 10.30158 | 10.360 | 10.585 | 10.3016 | 10.354 | 10.355 |
| 4S | 10.588 | 10.58000 | 10.580 | 11.148 | 10.580 | 10.572 | 10.579 |
| 1P | 9.608 | 10.02406 | 9.609 | 9.492 | 10.0246 | 9.891 | 9.899 |
| 2P | 10.110 | 10.30248 | 10.109 | 10.038 | 10.3029 | 10.258 | 10.260 |
| 1D | 9.841 | 10.30266 | 9.846 | 9.551 | 10.3032 | 10.156 | 10.164 |

Table3. Bottom-charmed $(b\bar{c})$ mass spectra in $GeV$ ($m_c = 1.209\ GeV, m_b = 4.823\ GeV$, $B = 0.202\ GeV^2$, $C = 1.213$, $\delta = 0.371\ GeV$)

| State | Present work | AEIM[11] | NU[15] | AIM[30] | LTM[18] | ANN[19] | Exp. [60] |
|---|---|---|---|---|---|---|---|
| 1S | 6.274 | 6.277473 | 6.277 | 6.277 | 6.2770 | 6.274 | 6.275 |
| 2S | 6.845 | 7.037641 | 7.383 | 6.814 | 7.0372 | 6.839 | 6.842 |
| 3S | 7.125 | 7.797808 | 7.206 | 7.351 | 7.7973 | 7.245 | |
| 4S | 7.283 | 7.038623 | | 7.889 | | 7.522 | |
| 1P | 6.519 | 7.798791 | 7.042 | 6.340 | 7.0381 | 6.743 | |
| 2P | 6.959 | | 6.663 | 6.851 | 7.7983 | 7.187 | |
| 1D | 6.813 | | | 6.452 | | 7.046 | |

Table4. Mass spectra of Charmed-strange $(c\bar{s})$ meson in $GeV$ ($m_c = 1.628\ GeV$, $m_s = 0.419\ GeV, B = 0.202\ GeV^2$, $C = 2.046, \delta = 0.561\ GeV$)

| State | Present work | NU[15] | AIM[30] | Exp. | |
|---|---|---|---|---|---|
| 1S | 1.969 | 1.968 | 2.512 | 1.96849 ± 0.00033 | [61] |
| 2S | 2.709 | 2.709 | 2.709 | 2.709 | [62] |
| 3S | 2.913 | 2.932 | 2.906 | | |
| 4S | 2.998 | | 3.102 | | |
| 1P | 2.601 | 2.565 | 2.649 | | |
| 2P | 2.876 | | 2.860 | | |
| 1D | 2.862 | 2.857 | 2.859 | 2.859 | [62] |

## 5. Conclusion

The energy eigenvalue equation of the SE with the KPIQP has been obtained with elegance. We demonstrated that the multiple turning point problems can be truncated to two turning points one for a non-symmetric potential function by using an appropriate Pekeris-type approximation scheme to the deal with the orbital centrifugal barrier energy. This approximation scheme allows us to solve the WKB integral analytically for any angular momentum quantum number. We computed the masses of mesons systems by using the Cornell potential as a model. The approximate mass spectra of the meson systems obtained in this present work are in excellent agreement with the results obtained by the other analytical methods and thus reinforce the exactness of the leading order WKB approximation method. Finally, our results may be useful to future experimental works and also can be applied in the study of molecular structures and interactions of diatomic molecules.

## Acknowledgment

It is a pleasure for us to thank the kind referees for their many useful comments and suggestions which greatly helped us in making improvements to this paper